\documentclass[articlepaper,english,reprint,nofootinbib,aps,superscriptaddress,showpacs,showkeys,prx]{revtex4-1}

\usepackage{babel,calc,amsmath,amsthm,amssymb,graphicx,subfigure,xcolor,comment}
\usepackage{txfonts}
\usepackage{mathdots}
\usepackage{todonotes}
\usepackage[T1]{fontenc}
\usepackage[unicode=true]{hyperref}
\usepackage{booktabs}
\usepackage{threeparttable}
\usepackage{times}
\usepackage{CJK}
\usepackage{bbold}
\usepackage{appendix}

\hypersetup{
	colorlinks=true,       		
	linkcolor=blue,          	
	citecolor=blue,            
	urlcolor=black,           	
}

\theoremstyle{definition}

\def\bra#1{\left\langle {#1} \right\rvert}
\def\ket#1{\left\lvert {#1} \right\rangle}
\def\braket#1{\left\langle {#1} \right\rangle}

\begin{document}


\title{Three-Photon Polarization Entanglement of Green Light}

\author{Yan-Chao Lou}
\email[]{These authors contributed equally to this work.}
\author{Zhi-Cheng Ren}
\email[]{These authors contributed equally to this work.}
\author{Chao Chen}
\email[]{chaochen@nju.edu.cn}
\author{Pei Wan}
\author{Wen-Zheng Zhu}
\author{Jing Wang}
\author{Shu-Tian Xue}
\author{Bo-Wen Dong}
\author{\\Jianping Ding}
\affiliation{National Laboratory of Solid State Microstructures and School of Physics, Nanjing University, Nanjing 210093, China}
\affiliation{Collaborative Innovation Center of Advanced Microstructures, Nanjing University, Nanjing 210093, China}

\author{Xi-Lin Wang\href{https://orcid.org/0000-0002-3990-6454}}
\email[]{xilinwang@nju.edu.cn}
\affiliation{National Laboratory of Solid State Microstructures and School of Physics, Nanjing University, Nanjing 210093, China}
\affiliation{Collaborative Innovation Center of Advanced Microstructures, Nanjing University, Nanjing 210093, China}
\affiliation{Hefei National Laboratory, Hefei 230088, China}
\affiliation{Synergetic Innovation Center of Quantum Information and Quantum Physics, University of Science and Technology of China, Hefei 230026, China}

\author{Hui-Tian Wang\href{https://orcid.org/0000-0002-2070-3446}}
\affiliation{National Laboratory of Solid State Microstructures and School of Physics, Nanjing University, Nanjing 210093, China}
\affiliation{Collaborative Innovation Center of Advanced Microstructures, Nanjing University, Nanjing 210093, China}
\affiliation{Collaborative Innovation Center of Extreme Optics, Shanxi University, Taiyuan 030006, China}

\date{\today}

\begin{abstract}
Recently, great progress has been made in the entanglement of multiple photons at various wavelengths and in different degrees of freedom for optical quantum information applied in diverse scenarios. However, multi-photon entanglement in the transmission window of green light under the water has not been reported yet. Here, by combining femtosecond laser based multi-photon entanglement and entanglement-maintaining frequency upconversion techniques, we successfully generate a green two-photon polarization-entangled Bell state and a green three-photon Greenberger-Horne-Zeilinger (GHZ) state, whose state fidelities are 0.893$\mathbf{\pm}$0.002 and 0.595$\mathbf{\pm}$0.023, respectively. Our result provides a scalable method to prepare green multi-photon entanglement, which may have wide applications in underwater quantum information.
\end{abstract}

\maketitle

\section{Introduction} 

\begin{figure}[b]
	\centering
	\includegraphics[width=0.95\linewidth]{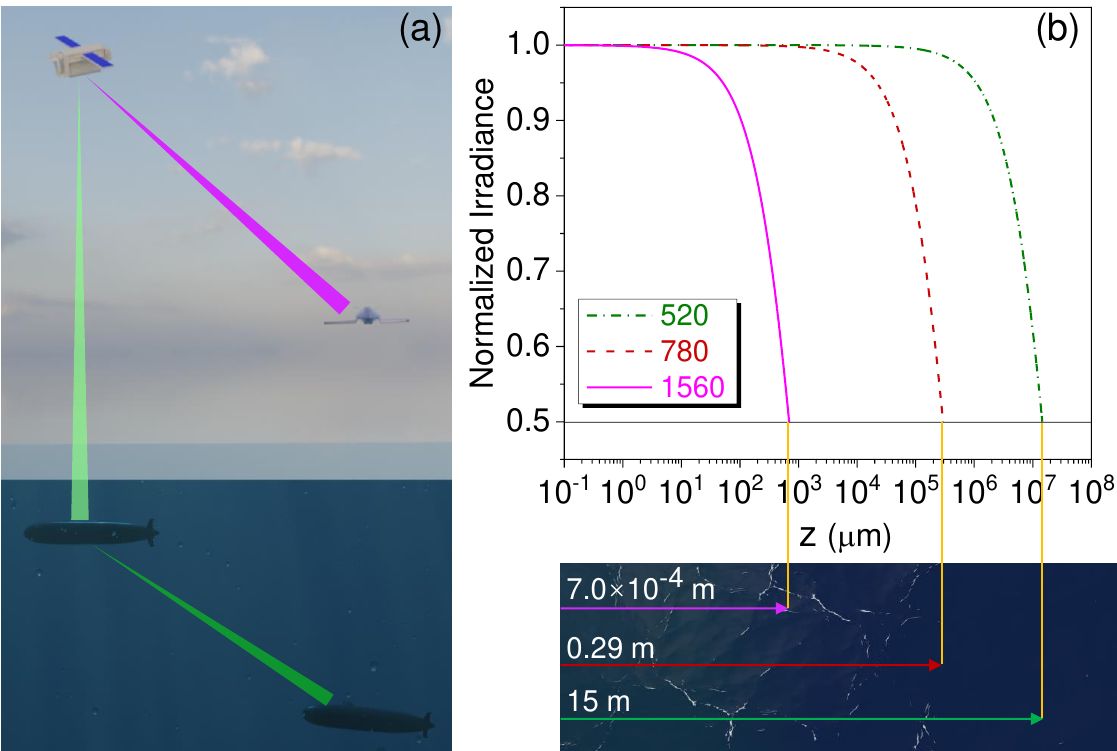}
	\caption{(a) Schematic scenarios of space-to-water and multi-party underwater quantum communication. (b) Propagation distance of light at wavelengths of 1560, 780 and 520 nm in water, when the irradiance is reduced to half. Photons at 520 nm have lower absorption and travel farther in water.}
	\label{fig: underwater comm}
\end{figure}

Multi-qubit entanglement is the fundamental quantum resource in quantum physics and quantum information applications~\cite{Einstein1935, Bell1964, Greenberger1989, bennett1995, Ekert1998}. Photon, due to the weak interaction with environment, is the most promising flying qubit, which can carrier quantum information in its various degrees of freedom, such as polarization and optical orbital angular momentum~\cite{Kwiat1995, Mair2001, Fickler2016}. Over the past two decades, multi-photon entanglement has played a crucial role in fundamental quantum physics and quantum information, such as in quantum teleportation~\cite{Bouwmeester1997, Murao1999, wang2015, Ren2017}, the test of the quantum nonlocality~\cite{freedman1972, Pan2000, hensen2015} and long-distance quantum communication~\cite{Ma2012, Juan2017, Yin2020}. Compared with quantum communication based on mobile devices such as satellites and drones over the continent~\cite{Liu2021, liao2018, Chen2021}, space-to-water and underwater quantum communication shown in Fig.~\ref{fig: underwater comm}(a) is so far lagging behind. Because of the absence of underwater entangled photon sources, previous underwater quantum communications were all implemented with single photons~\cite{Hu2019, Ji2017, Bouchard2018}. As is well known, the absorption of light at different wavelengths varies exceedingly in water, as shown in Fig.~\ref{fig: underwater comm}(b)~\cite{Raymond1981, Deng2012}. Green light lies in the transmission window of light in water~\cite{schirripa2020}, which makes green-photon entanglement essential for underwater and space-to-water quantum communication.

Although spontaneous parametric down conversion (SPDC) has been widely used to prepare multi-photon entanglement~\cite{Lu2007, Yao2012, Zhang2015, Wang2016}, including the record 12-photon entanglement at 1560 nm~\cite{Zhong2018}, generation of green multi-photon entanglement states has long been a challenge. Using the SPDC from 260 nm to 520 nm to directly prepare green-photon entanglement faces three knots. One is that few high-power laser is available around 260 nm to pump the required SPDC process. Second, appropriate nonlinear crystals are relatively absent, because the known nonlinear crystals, such as lithium niobate (LN) and potassium titanyl phosphate (KTP), have generally strong absorption in ultraviolet and deep ultraviolet regimes~\cite{Dmitriev2013} or such as $\beta$-barium borate ($\beta$-BBO) and lithium triborate (LBO), are very difficult to achieve the phase matching although they are transparent in those regimes. Third, commercial optical components that can be used for 260 nm are also scarce.

The mainstream method for obtaining short-wavelength quantum sources relies on single photon frequency conversion~\cite{kumar1990, Raymer2012, Kuzucu2008, Vollmer2014, Zhou2016, Zhou2017, Liu2020, Tyumenev2022}, which has shown great potential in connecting platforms working on different wave bands~\cite{Tanzilli2005, Lauk2020, Rakher2010}. According to the interaction distance between photons and nonlinear materials resulting into different conversion efficiency defined by the ratio between the converted photon number and the incoming number of signal photons, frequency conversion could be divided into two types with long and short interaction distances. To achieve long interaction distance, one can choose interacting light with long coherence length including continuous light and picosecond (ps) pulsed laser, which supports the nonlinear interaction in a long medium including a waveguide to achieve a high upconversion efficiency of up to tens of percents~\cite{Rakher2010}. However, on some special application scenarios, for example, to realize spatial structure preserving frequency conversion, the interaction is restricted near the image plan~\cite{Li2013, Yang2022} and the corresponding interaction length is limited. In these circumstances, femtosecond (fs) pulsed laser with a high peak power would be a better choice although the conversion efficiency is several percents~\cite{Qiu2023, Sephton2023}, about one order lower than that with ps pulsed laser. 

So far, the most widely used approach to prepare multi-photon entanglement source is to connect multiple pairs of entangled two photons~\cite{Pan2012}, which requires high indistinguishability between independent photons from different pairs in temporal, frequency and spatial domains. To achieve high indistinguishability in time domain, pulsed light are employed as pump source. More specifically, the fs pulse is much easier to realize high indistinguishability than the ps pulse that need strongly filtering. Therefore, the brightness of multi-photon source based on the fs pulse is much higher than the ps pulse. For example, for a  4-photon entanglement source by connecting two pairs of entangled two photons, the brightness with the fs pulse is about 5 orders higher than the ps pulse~\cite{Wang2016, Zhong2018, Bao2023, Valivarthi2016, Sun2016}. Up to now, the maximum entangled photons with the ps pulse is 4~\cite{Bao2023}, while the fs pulse has enabled successful demonstration of up to 12-photon entanglement~\cite{Zhong2018}. Therefore, the fs pulse is more suitable for achieving short-wavelength multi-photon entanglement source. 

Here, we design a protocol to generate three-photon entanglement of green light, by firstly preparing multi-photon (include infrared and green photons) polarization-entangled state vis the fs laser pumped SPDC and then exploiting frequency upconversion to convert the infrared photons to green ones. To realize the frequency conversion of a photon that is polarization entangled with other photons, we imbed the frequency transducer in a Sagnac interferometer~\cite{Ren2021, Lou2022}, where the horizontally and vertically polarized components are both frequency upconverted by passing through the transducer in opposite directions. After outputting from the Sagnac interferometer, orthogonally polarized upconverted photons coherently combine and maintain their polarization entanglement with other photons~\cite{ramelow2012}. Until now, it remains a challenge to realize the fs pulse-based polarization entanglement-maintaining frequency upconversion.

\section{Experimental Details} 

\begin{figure}[!ht]
	\centering
	\includegraphics[width=\linewidth]{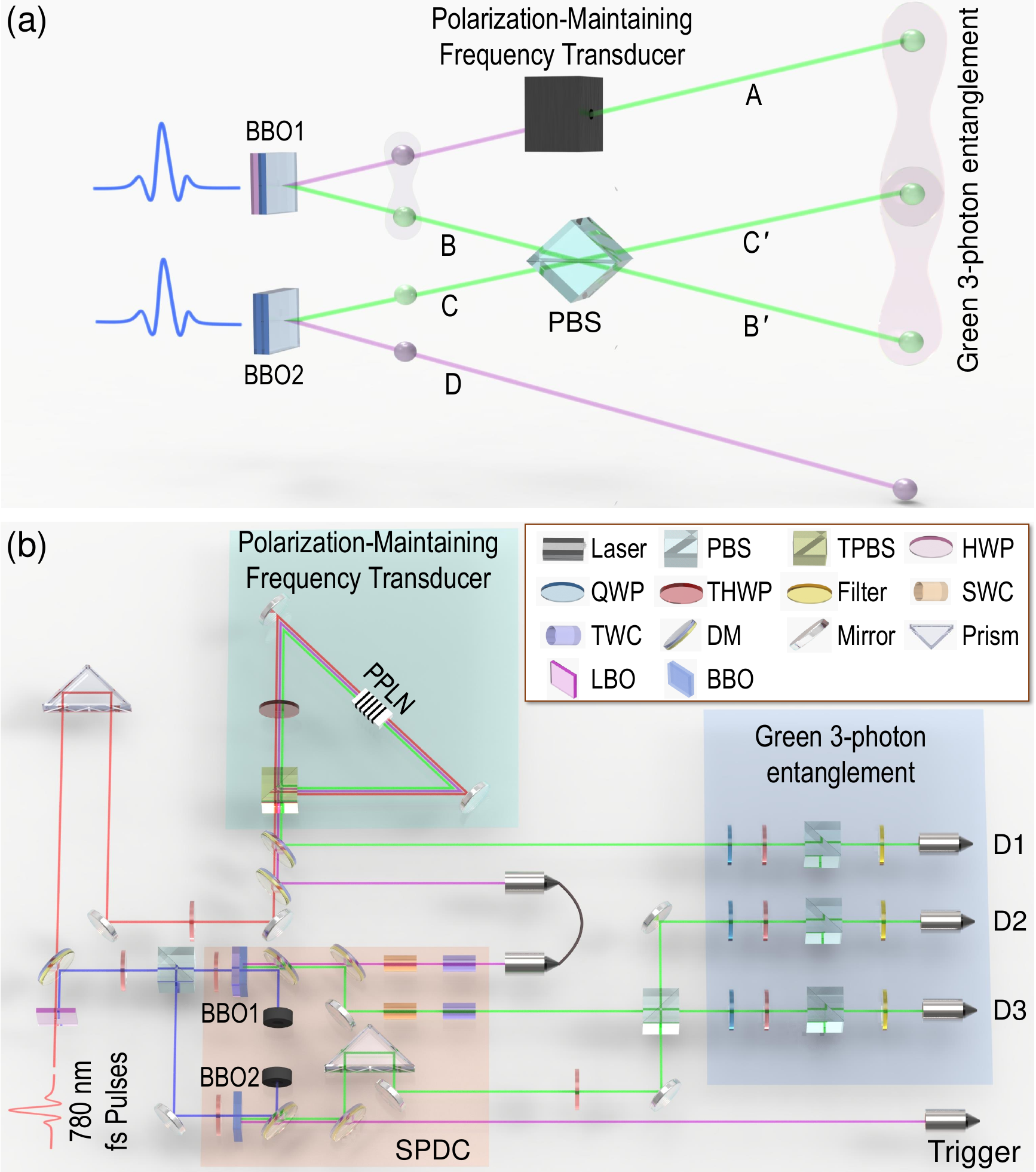}
	\caption{(a) Schematic diagram for preparing a green three-photon entanglement state. (b) Experimental setup. A fs pulsed laser at 780 nm is focused on a frequency-doubling LBO crystal to generate fs laser pulses at 390 nm, which is the pumps of the two crystals (BBO1 and BBO2) where pairs of single photons are generated. The PPLN is used for upconverting the single photon at 1560 nm to 520 nm under the pump of another laser at 780 nm. Due to the Sagnac interferometer configuration, the frequency upconversion is polarization independent, with the coherence in the polarization state being preserved. The triple-polarization beam splitter (TPBS) and triple half-wave plate (THWP) in the Sagnac interferometer function at three wavelengths of 1560, 780 and 520 nm. The three green photons are finally detected by detectors D1, D2 and D3, heralded by a single photon at 1560 nm, which is coupled into detector D4.}
	\label{fig:exp_setup}
\end{figure}

In this article, we develop a polarization-entanglement-maintaining Sagnac frequency transducer technology to get 520-nm three-photon entanglement based on the SPDC sources as shown in  Fig.~\ref{fig:exp_setup}(a). First, a pair of entangled photons at 1560 and 520 nm is generated by the SPDC pumped by a fs laser at 390 nm that is a frequency doubling of the fs laser at 780 nm. Then polarization-entanglement-maintaining frequency upconversion is implemented to transfer the frequency of the single photon from 1560 nm to 520 nm. The scalability of this scheme pumped by the fs laser is shown by interfering one of the entangled two green photons with another photon pair (at 520 and 1560 nm) to generate three-photon entanglement at 520 nm heralded by a single photon at 1560 nm. Our green three-photon entanglement combines the fs laser pumped SPDC with the frequency upconversion technique, which provides an innovative way to generate scalable entanglement of green photons, and may pave the way for entanglement based underwater quantum communication.

As the experimental setup shown in Fig.~\ref{fig:exp_setup}(b), the 140-fs pulsed laser at 780 nm (Chameleon, Coherent) is divided into two beams. One beam is frequency-doubled into 390 nm in a LBO crystal; the other serves as the pump laser for upconverting a photon from 1560 nm to 520 nm. The fs laser at 390 nm is also divided and focused to pump BBO1 and BBO2 individually, with the same waist radii of 30 $\mu$m. Notably, BBO1 consists of two glued BBO pieces ($7 \times 7 \times 2$ mm$^3$, the cutting angles are $\theta=30.5^{\circ}$ and $\phi=30^{\circ}$) with orthogonal optical axes~\cite{Kwiat1999}. A $45^\circ$-polarized fs pump laser at 390 nm is incident into BBO1, type-II non-degenerate SPDC could happen either in the first BBO piece or in the second piece. SPDC that occurs in the two glued BBO pieces will introduce walk off. After the space walk-off compensation (SWC) and time walk-off compensation (TWC) operations~\cite{Kim2001, Kim2000}, the down-converted photon pairs from the two BBO pieces are superposed to be a polarization entangled state, i.e., $\frac{1}{\sqrt{2}} ( | H \rangle_{520} | V \rangle_{1560} + | V \rangle_{520} | H \rangle_{1560} )$. The possible phase between $|H\rangle_{520} |V\rangle_{1560}$ and $|V\rangle_{520}|H\rangle_{1560}$ can be eliminated by introducing a Babinet compensator in the path of 520 nm photon (See Appendix A for details). Using dichroic mirrors (DMs), we filter out the pump laser and separate the photons at 1560 nm and 520 nm into two paths. Before coupling a single photon at 1560 nm into an entanglement-maintaining frequency transducer, we added a filter with a full width at half maximum (FWHM) of 25 nm to remove the stray light. Another filter with a FWHM of 3 nm is added for the single photon at 520 nm before it is detected.

A Sagnac interferometer is employed to achieve the polarization-entanglement-maintaining frequency transducer. The single photon at 1560 nm is upconverted to 520 nm at a periodically poled lithium niobate (PPLN) crystal ($1 \times 2 \times 3$ mm$^3$) in the type-0 phase matching configuration, pumped by a fs pulsed laser at 780 nm with a power of 850 mW. The photons at 1560 nm in the Sagnac interferometer has a focal radius of 160 $\mu$m measured by a camera (SP907-1550, Ophir Spiricon Inc.). We design the poled period (6.97 $\mu$m) of the PPLN crystal to maximize the conversion efficiency. Moreover, the crystal is maintained at the optimal phase-matching temperature of $135^{\circ}$C. With the help of the PBS and a half-wave plate (HWP) in the Sagnac interferometer, the horizontally and vertically polarized input photons at 1560 nm pass through the PPLN crystal from the opposite directions, and are upconverted to green photons at 520 nm of vertical and horizontal polarizations, respectively. The entangled states in the Sagnac interferometer experience the evolutions as follows
\begin{equation}
\begin{aligned}
{\rm Clockwise \circlearrowright:} & \quad | H \rangle_{1560} \xrightarrow{\rm HWP} | V \rangle_{1560} \xrightarrow{\rm PPLN} | V \rangle_{520} \\
{\rm Anticlockwise \circlearrowleft:} & \quad | V \rangle_{1560} \xrightarrow{\rm PPLN} | V \rangle_{520} \xrightarrow{\rm HWP} | H \rangle_{520}.
\end{aligned}
\end{equation}
Clearly, when the photon's frequency is upconverted, its polarization is flipped. Therefore, the original entangled state is transformed as
\begin{equation}
\begin{aligned}
\frac{1}{\sqrt{2}} ( | H \rangle_{520} | V \rangle_{1560} + | V \rangle_{520} | H \rangle_{1560} ) \\ \xrightarrow{\rm Transducer} \frac{1}{\sqrt{2}} ( | H \rangle_{520} | H \rangle_{520} + | V \rangle_{520} | V \rangle_{520} ).
\label{eq:bell}
\end{aligned}
\end{equation}

Another horizontally polarized laser beam at 390 nm pumps the BBO2, where a photon pair with the polarization state of $| V \rangle_{520} | H \rangle_{1560}$ is generated by the SPDC. After DMs, the pump laser at 390 nm is filtered out, and the photons at 1560 and 520 nm are separated. The single photon at 1560 nm is directly coupled into the single-mode fiber, and measured by the superconducting nanowire single-photon detector (P-SPD-32S, Photon Technology Co.,Ltd.). As for the photon at 520 nm, it is prepared in the polarization state of $\frac{1}{\sqrt{2}} ( | H \rangle_{520} + | V \rangle_{520} )$, and then interferes with one of the two entangled green photons at a PBS. By moving the prism placed on a translation stage, the two photons for interference arrive at the PBS simultaneously. We post-select the events that coincidence counts are measured at the two output ports of the PBS, and finally successfully generate the three-photon entanglement state at 520 nm written as 
\begin{align}
\frac{1}{\sqrt{2}} ( | H \rangle_{520} | H \rangle_{520} | H \rangle_{520} + | V \rangle_{520} | V \rangle_{520} | V \rangle_{520} ),
\label{Eq3}
\end{align}
which is accompanied by a single photon at 1560 nm. We provide details about the post-selected interference at PBS in Appendix B.

In the experiment, to obtain entangled photon pair, we need to compensate for both the time and space walk off between horizontally and vertically polarized photons at 1560 nm (520 nm) downconverted in the first and second crystal pieces in BBO1, respectively. After several experimental attempts, the compensation scheme is determined. For the SWC of the photon at 1560 nm, two quartz crystals ($\theta=45^{\circ}$) are added. One has a length of 20.14 mm and its optical axis is along the horizontal direction, while the other has a length of 25.48 mm and its optical axis is along the vertical direction. The TWC of photon at 1560 nm is accomplished by introducing a quartz crystal ($\theta=0^{\circ}$) with a length of 16.59 mm and its optical axis parallel to the horizontal direction. Because of the dispersion of BBO, the compensations for photons at 1560 nm and 520 nm are different. For the SWC of the photon at 520 nm, we use two quartz crystals ($\theta=45^{\circ}$) with the same length of 13.37 mm and whose optical axes along orthogonal directions, i.e., horizontal and vertical directions, respectively. The TWC of the photon at 520 nm is achieved by a quartz crystal ($\theta=0^{\circ}$) with a length of 15.75 mm whose optical axis is along the horizontal direction. 

Another vital improvement is that we developed the fs laser pumped single-photon frequency upconversion technique. In comparison, previous related experiments were implemented in a continuous-wave domain, which is not scalable in generating multi-photon entanglement, limiting its application in optical quantum computing and information. The main challenge of using the fs pump laser is to achieve the temporal alignment between the 1560-nm single photon to be upconverted and the 140-fs pump laser at 780 nm in the frequency transducer---the PPLN crystal. Here, we first block the pump laser and couple the single photon at 1560 nm into the Sagnac interferometer. At the output of the interferometer, the single photon at 1560 nm is measured by a single photon detector, and its arrival time is recorded by a time-digital converter with a digital resolution of up to 1 ps (Time Tagger Ultra, Swabian Instruments Inc). Then we blocked the single photon at 1560 nm and turned on the pump laser at 780 nm. Although the PPLN crystal is not designed for parametric downconversion of photons from 780 nm to 1560 nm, a weak signal at 1560 nm can be still generated under a strong pump. The arrival time at the detectors of photons (1560 nm) spontaneously downconverted from the pump laser is also measured and tuned to be consistent with that of the input single photon at 1560 nm by moving the delay prism in the pump laser (780 nm) path. We finely tune the position of the prism to optimize upconversion efficiency and to obtain the brightest single photon at 520 nm.

\section{Results and Dicussions}

\begin{figure}[b]
	\centering
	\includegraphics[width=0.95\linewidth]{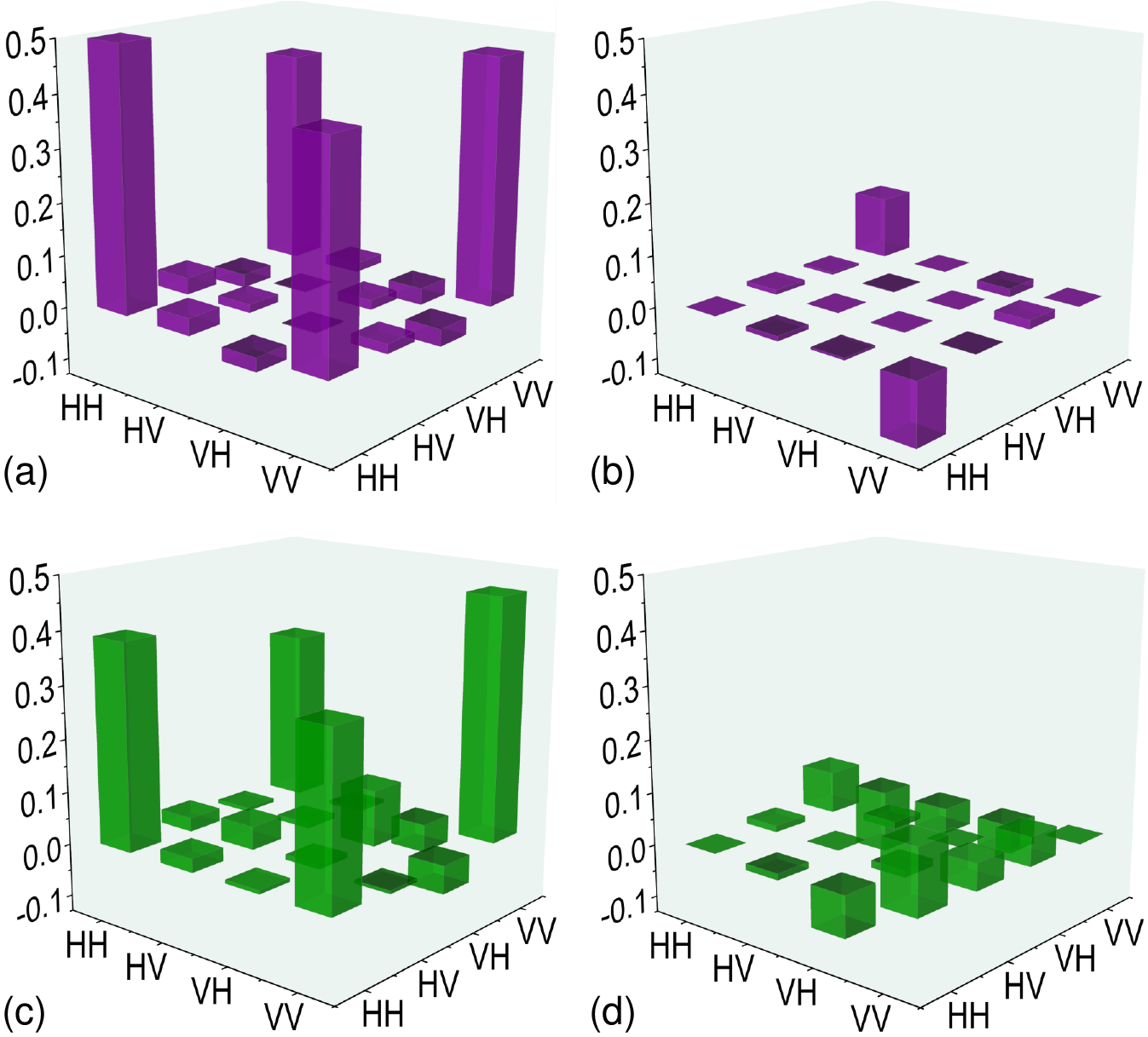}
	\caption{Reconstructed density matrix of two-photon entangled state. (a) and (b) The measured real and imaginary parts of the density matrix of the generated quantum state, whose target state is $\frac{1}{\sqrt{2}} ( | H \rangle_{520} | H \rangle_{1560} + | V \rangle_{520} | V \rangle_{1560} )$.  (c) and (d) The real and imaginary parts of the density matrix of the quantum state, with the target state of $\frac{1}{\sqrt{2}} ( | H \rangle_{520} | H \rangle_{520} + | V \rangle_{520} | V \rangle_{520} )$.}
	\label{fig3}
\end{figure}
Under our experimental conditions, the produce efficiency of degenerate two-photon pair at 520 nm is about 0.25\% with respect to the two-photon pair at 520 nm and 1560 nm, which can be promoted by increasing the power density of the pump laser. We measure the entangled two-photon states before and after the frequency upconversion by quantum state tomography~\cite{James2001}, requiring that each photon is measured in $\ket{H}$, $\ket{V}$, $\frac{1}{\sqrt{2}} ( \ket{H} +\ket{V} )$, $\frac{1}{\sqrt{2}} ( \ket{H} -\ket{V} )$, $\frac{1}{\sqrt{2}} ( \ket{H} + i\ket{V} )$ and $\frac{1}{\sqrt{2}} ( \ket{H} -i\ket{V} )$ basis. For measuring the entangled state before the frequency upconversion, we flip the polarization of the single photon at 1560 nm to have the same polarization entanglement state as that after the frequency conversion, i.e., $\frac{1}{\sqrt{2}} ( \ket{HH} + \ket{VV} )$. The reconstructed density matrices are shown in Fig.~\ref{fig3}. The fidelity of measured states before and after the frequency upconversion are 0.893$\pm$0.002 and 0.746$\pm$0.011, confirming that the polarization entanglement is preserved during the frequency upconversion. We note that a weak signal at 520 nm is generated spontaneously under the pump of the laser at 780 nm. This background may lead to a decrease in the state fidelity after frequency upconversion. To further promote the state fidelity, we can decrease the pump power and increase the detection efficiency to lower the impact of higher orders photon-pair events~\cite{Wang2016}.

\begin{figure}[!b]
	\centering
	\includegraphics[width=0.90\linewidth]{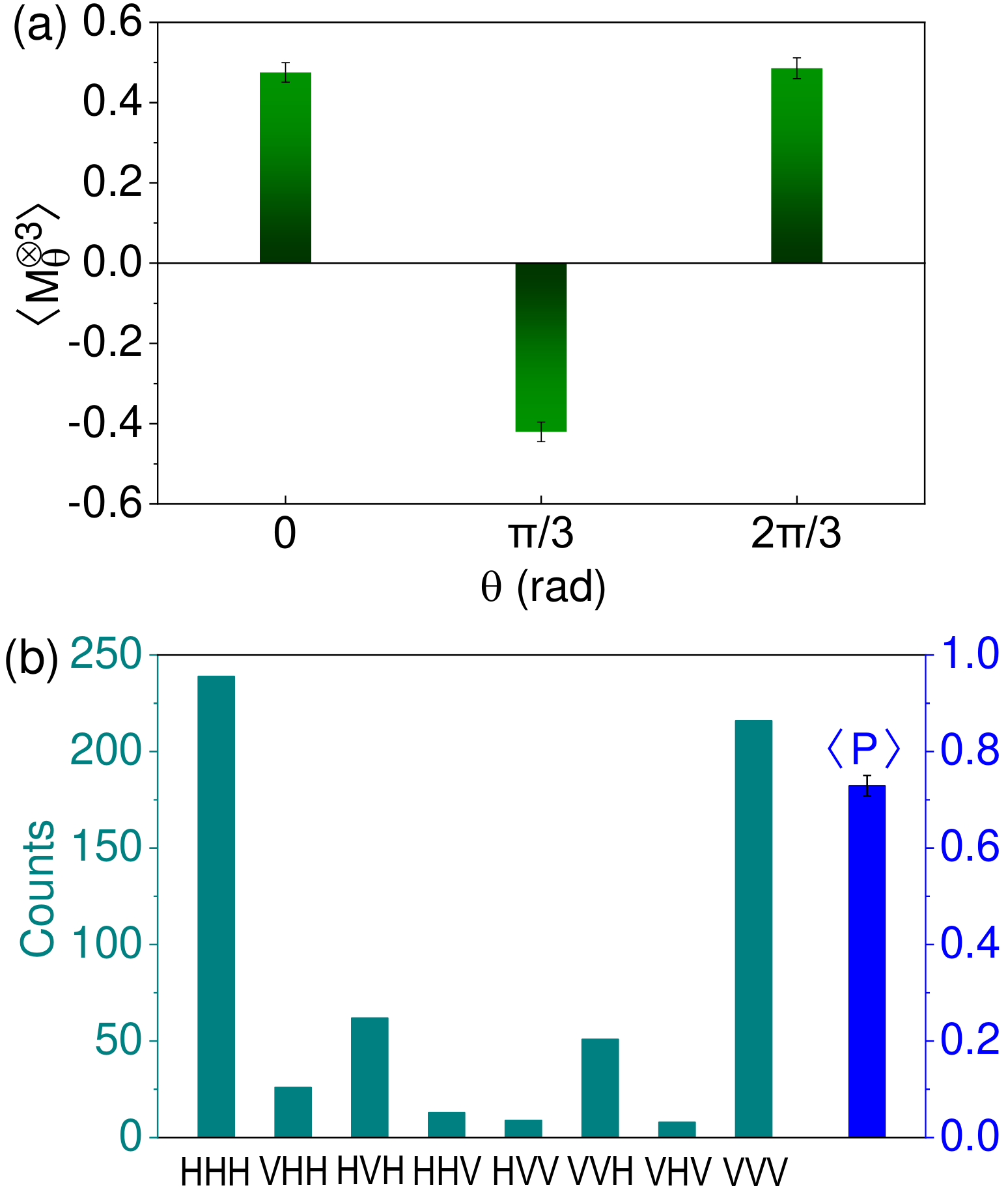}
	\caption{Experiment results of green three-photon polarization entanglement. (a) Expectation values of $\braket{M_{k\pi/3}^{\otimes 3}}$ measured in the basis of $\frac{1}{\sqrt{2}} ( | H \rangle \pm e^{ik \pi /3} | V \rangle)$. (b) Coincidence counts when measuring the three photons in $\ket{H/V}$ basis. The calculated $\braket{P}$ is shown in the right.}
	\label{fig4}
\end{figure}
Benefiting from the advantage of using fs pulsed laser in our experiment, the entangled photon number can be scaled by interfering with one of the entangled green photons with extra green photon. We now have another photon pair in the state of $\ket{H}_{1560}\ket{V}_{520}$ generated from the BBO2 crystal. The single photon at 520 nm is prepared in polarization direction of $45^\circ$ and then interferes with one of the entangled two green photons at PBS. Heralded by the single photon at 1560 nm, we post-select the coincidence at the two output ports of the PBS, together with the upconverted green photon, which is proved to be a green three-photon GHZ state. 

The prepared GHZ state is analyzed by the following measurement. The coherence of the three-photon GHZ state, a reflection of the off-diagonal element of its density matrix, is weighted by
\begin{equation}
C = \frac{1}{3}\sum_{k=0}^2(-1)^k \braket{M_{k\pi/3}^{\otimes 3}},
\end{equation}
where $M_{k\pi/3}=\cos{(k\pi/3)}\sigma_x + \sin{(k\pi/3)}\sigma_y,~k=0,1,2$. The expectation of observable $M_{\theta}^{\otimes 3}$ can be obtained by measuring the three photons individually in the basis of $\frac{1}{\sqrt{2}} ( \ket{H} \pm e^{ i \theta}\ket{V} )$, where $\theta\in \{0, \pi/3, 2\pi/3\}$. We also measure the state in the $\ket{H}/\ket{V}$ basis to get the diagonal element information of the density matrix, denoted by $P = \ket{HHH}\bra{HHH} + \ket{VVV}\bra{VVV}$. It is proved that the fidelity of generated state with the perfect three-photon GHZ state as shown in Eq.~(\ref{Eq3}) can be calculated by 
\begin{equation}
F = \dfrac{1}{2} (\braket{P} + \braket{C}),
\end{equation}
where $F$ represents the state fidelity~\cite{Wang2016, Bourennane2004}. 

In Fig.~\ref{fig4}, we show the measured expectation values of observables $M_\theta^{\otimes{3}}$ and $P$ for the three-photon state, which are collected in the time duration of one hour. The errors are estimated under the assumption that the photon statistics follow a Poisson distribution. From the measured results, the state fidelity we calculated is $0.595\pm0.023$. To estimate the entanglement of the prepared state, we specify the entanglement witness as $W=\alpha I +\ket{\psi}\bra{\psi}$, where $I$ is an identity matrix, $\ket{\psi}$ is the desired entanglement state as shown in Eq.~(\ref{Eq3}), and $\alpha$ is 0.5 for the GHZ states ~\cite{Bourennane2004,guhne2007}. It can be found that entanglement witness is related to the state fidelity by $\braket{W}=0.5+F$. To signal the entanglement, the expectation value of $W$ should be negative; equivalently, the state fidelity of the GHZ state must exceed 0.5. The fidelity we measured for three-photon GHZ state exceeds the entanglement threshold 0.5 by 4.13 standard deviations, implying the presence of genuine entanglement of the three photons.

The experimental imperfections are mainly ascribed to the limited precision of the time and space walk-off compensation and the nonunity of the frequency upconversion. For instance, the fidelities of the original two-photon entangled states (at 1560 and 520 nm) can be further promoted by choosing the compensation quartz crystals with accurate thickness. To ensure that all single photons have the probability of being frequency upconverted, we should match the pump laser with the single photon to be converted in all degrees of freedom. In our experiment, the bandwidth of the single photon is wider than the effective bandwidth for the frequency upconversion process. To further increase the upconversion efficiency, one can shape the spectra of the single photon. Otherwise, the key method to improve the upconversion efficiency is to raise the pump power. 

\section{Conclusions} 

We have developed a polarization-entanglement-maintaining frequency transducer technology of fs pump and experimentally achieved two- and three-green-photon entanglement. Due to the high transmission of green photons in the water, our results not only pave the way for low-loss underwater and air-to-water multi-party quantum information processes~\cite{Hu2019, Bouchard2018}, but also provide opportunity to address the issues of heat dissipation for future quantum machines by placing them in underwater environment on seabed. Furthermore, the frequency transducer technology of fs pump can be used to realize the polarization-entanglement-maintaining upconversion with ultrathin nonlinear materials~\cite{Abdelwahab2022, Rocio2021}, and the generated multi-photon entangled source with short wavelength may have applications in quantum metrology~\cite{Giovannetti2011} including supperresolving phase measurement~\cite{Zhou2017}. 

\begin{acknowledgments}
This work was supported financially by the National Natural Science Foundation of China (Grants No. 12234009, No.12275048, No. 12304359, and No. 12274215); the National Key R\&D Program of China (Grants No. 2019YFA0308700 and No. 2020YFA0309500); the Innovation Program for Quantum Science and Technology (Grant No. 2021ZD0301400); the Program for Innovative Talents and Entrepreneurs in Jiangsu; the Natural Science Foundation of Jiangsu Province (Grant No. BK20220759); the Key R\&D Program of Guangdong Province (Grant No. 2020B0303010001); the China Postdoctoral Science Foundation (2023M731611); the Jiangsu Funding Program for Excellent Postdoctoral Talent (2023ZB717).
\end{acknowledgments}

\appendix

\section{COMPENSATION OF THE PHASE IN THE ENTANGLED STATES}
There could be an arbitrary phase $\phi$ between the term $\ket{H}_{520}\ket{V}_{1560}$ and $\ket{V}_{520}\ket{H}_{1560}$, then the state on the left of the arrow in Eq.~\ref{eq:bell} becomes
\begin{equation}
    \frac{1}{\sqrt{2}}(\ket{H}_{520}\ket{V}_{1560}+e^{\mathrm{i}\phi}\ket{V}_{520}\ket{H}_{1560}).
    \label{eq:phase1}
\end{equation}
To prepare desired green-photon Bell state shown on the right of the arrow in Eq.~(\ref{eq:bell}), we eliminate the phase in Eq.~(\ref{eq:phase1}) by using a Babinet compensator, which introduces a variable phase between the horizontally and vertically polarized component for a single photon. 

Rephrasing the state Eq.~(\ref{eq:phase1}) in basis vectors of $\ket{A}$ and $\ket{D}$, we get
\begin{equation}
\begin{aligned}
   \frac{1}{2\sqrt{2}} \left[  (1+e^{\mathrm{i}\phi})\ket{D}_{520}\ket{D}_{1560}-(1-e^{\mathrm{i}\phi})\ket{D}_{520}\ket{A}_{1560} \right. 
    \\
   \left. +(1-e^{\mathrm{i}\phi})\ket{A}_{520}
\ket{D}_{1560}-(1+e^{\mathrm{i}\phi})\ket{A}_{520}\ket{A}_{1560} \right],\\
\label{eq:phasead}
\end{aligned}
\end{equation}
where $\ket{A}=\frac{1}{\sqrt{2}}(\ket{H}+\ket{V})$ and $\ket{D}=\frac{1}{\sqrt{2}}(\ket{H}-\ket{V})$. If we project the state Eq.~(\ref{eq:phasead}) onto $\ket{D}_{520}\ket{A}_{1560}$ or $\ket{A}_{520}\ket{D}_{1560}$, the projection probability $P$ is associated with the phase $\phi$ via $P=\frac{1}{4}(1-\cos{\phi})$. When $\phi=0$, the probability of projecting onto state $\ket{D}_{520}\ket{A}_{1560}$ or $\ket{A}_{520}\ket{D}_{1560}$ vanishes. In the experiment, we project the two photons onto the $\ket{D}$ and $\ket{A}$  basis respectively, and adjust the Babinet compensator to get the minimum two-fold coincidence, and then the unknown and arbitrary phase is eliminated. The phase could appear between the two terms in Eq.~(\ref{Eq3}) is also eliminated in the same way.

\section{ENTANGLING PHOTONS BY POST-SELECTED INTERFERENCE}
After the frequency transducer, the two entangled photons of 520 nm in paths $A$ and $B$ are in the state of
\begin{equation}
    \frac{1}{\sqrt{2}}(\ket{H}^A_{520}\ket{H}^B_{520}+\ket{V}^A_{520}\ket{V}^B_{520}).
\end{equation}
As shown in Fig.~\ref{fig:exp_setup}, to entangle the Bell state with the heralded single photon of diagonal polarization in path $C$, we couple the photon in paths $B$ and $C$ into the two ports of a PBS. Because the PBS transmits horizontally polarized photons and reflects vertically polarized photons, the PBS transform the input state as
\begin{equation}
\begin{aligned}
   \frac{1}{2}(\ket{H}^A_{520}\ket{H}^B_{520}+\ket{V}^A_{520}\ket{V}^B_{520})\otimes(\ket{H}^C_{520}+\ket{V}^C_{520})
   \\ \xrightarrow{\rm PBS}  \frac{1}{2}(\ket{H}^A_{520}\ket{H}^{B'}_{520}+\ket{V}^A_{520}\ket{V}^{C'}_{520})\otimes(\ket{H}^{C'}_{520}+\ket{V}^{B'}_{520}).
\end{aligned}
\end{equation}
We post-selected the events that cause coincidence counts of detectors in all the paths $A$, $B'$ and $C'$, then the state becomes
\begin{equation}
    \frac{1}{\sqrt{2}}(\ket{H}^A_{520}\ket{H}^{B'}_{520}\ket{H}^{C'}_{520}+\ket{V}^A_{520}\ket{V}^{B'}_{520}\ket{V}^{C'}_{520}),
\end{equation}
which is a three-photon GHZ state that is the same as Eq.~(\ref{Eq3}). 


\end{document}